\documentclass{elsart}

\usepackage{epsfig}

\begin{document}
\begin{frontmatter}




\title{Speed-up through entanglement -- many-body effects in neutrino processes}  


\author[a1]{Nicole F. Bell}\ead{nfb@fnal.gov},
\author[a2]{Andrew A. Rawlinson}\ead{a.rawlinson@physics.unimelb.edu.au} \, and
\author[a3]{R. F. Sawyer}\ead{sawyer@vulcan.physics.ucsb.edu}

\address[a1]{NASA/Fermilab Astrophysics Center, Fermi National Accelerator 
Laboratory, Batavia, Illinois 60510-0500, USA}
\address[a2]{School of Physics, The University of Melbourne,  Victoria 3010, 
Australia}
\address[a3]{Department of Physics, University of California at Santa Barbara,
Santa Barbara, California 93106, USA}

\begin{abstract}
We study a system containing many particles of identical kinematics
with a zero range interaction that scatters one from the other, and with the possible
exchange of an attribute. Taking an initial condition in which the attribute
is asymmetrically distributed in the regions of momentum
space occupied by the particles, we study the rate at which it becomes uniformly
distributed, through collisions. We find, in some circumstances, a
rate that is much faster than that which would be 
estimated from cross-sections. This behavior is attributable in some 
general sense to N-particle entanglement. We suggest applications to neutrino
physics, where the attribute is neutrino flavor.
\end{abstract}


\end{frontmatter}

\vspace{-0.5cm}
{\small FERMILAB-Pub-03/062-A}
\vspace{-0.5cm}

\section{Introduction}

We consider a system consisting of stable or quite long-lived particles (i.e.
stable for the time range that in which we shall be interested), confined in a
box for simplicity, and interacting occasionally with one another.  Taking an
initial state which is, at least with respect to some of its attributes, not in
statistical equilibrium, we can discuss time scales for evolution of the gross
features of the system.  In general these scales are determined by
cross-sections. But an exception to this assertion about scales and
cross-sections can be found, among other places, in the transformation of one
state of a particle into another through  evolution that can occur in isolation
from other particles, such as in neutrino oscillations or the precession of
spins in external magnetic fields. In this case it may not be to the point to
discuss the instantaneous``rates" at which the system changes, since we are
watching coherent development of an attribute of the system with time
behavior generically like $\sin \omega t$.

In the present note we demonstrate some circumstances under which we do
\underline{not} have such slow single-particle ``precessions" but in which
there is a long term coherent process arising from the short range interactions
of particles, which can be randomly distributed in momentum space. This
behavior can arise when the particles have another attribute that can be traded
in the course of an interaction. Both to have a concrete framework and with a
view to a possible application, we shall consider a system of neutrinos.  In
order to eliminate extraneous effects, we consider an example in which there
are two flavors of massless neutrinos, designated $\nu_e$ and $ \nu_{\tau}$,
and there is \underline{no} neutrino flavor mixing  in the Hamiltonian. There
is an initial distribution of the neutrinos specified, one in which there is
some systematic flavor asymmetry, such as neutrinos of one flavor being
predominantly of higher energy than those of the other flavor, or with
different angular distributions for the two species. The question we pose is:
``At what later time would the distributions would become more or less equal
through neutral current interactions?"  The conventional answer, as mentioned
above, is that this time is determined by a scattering rate proportional to the
weak cross-section times the density of scatterers. Equations in the literature
do indeed  predict shorter time scales for effects arising from
neutrino-neutrino interactions  in the case in which there is flavor
oscillation built into the Hamiltonian \cite{pantaleone}. The effects studied  in
these papers are analogous to the ``forward scattering", or ``index of
refraction" terms familiar in the study of the passage of neutrinos through
matter that contains electrons, in that they have an inverse time scale that is
proportional to $G_F \rho$ where $\rho$ is the neutrino number-density of the
medium and $G_F$ is the Fermi constant. This is in contrast to cross-section
effects, which are of order $G_F^2 \rho\,\omega^2$, where $\omega$ is the
energy of the scattering particles. 
These studies, however, predict that such effects are
\underline{strictly absent}, for pure neutrino systems, in the absence of 
flavor mixing in the
Hamiltonian, if one one begins, as we do, with a state that is diagonal 
in the flavor space \cite{raff}.
The essential difference between our approach and that of previous work is that
we retain much of the full complexity of the multi-body physics involved,
rather than assuming we may describe our ensemble of particles with
a single body density matrix.\footnote{Some of the subtleties involved in 
neutrino-neutrino forward scattering have recently been re-examined by 
A. Friedland and C. Lunardini \cite{fl}.  In particular, they raise the point 
that there may be circumstances in which a single body description is
inadequate, as is certainly the case for the effects we study here.
However, we do not see how to obtain the results of the present paper using
the perturbative approach of these authors.}

\section{Model}

The
basic process that we consider is simply 
\begin{equation}
\nu_e({\bf p})+ \nu_\tau({\bf q}) \rightarrow \nu_e({\bf q})+ \nu_\tau({\bf p}),
\label{reaction}
\end{equation} 
where the momenta ${\bf p}$ and ${\bf q}$ are drawn from the initial distributions.
These processes will tend to reduce the correlations of flavor and momentum 
that
we assume are present in the initial state. The question is: ``At what 
rate?"
The part of the Hamiltonian that will provide our effects operates only in the
subspace of the initial momentum states. We define annihilation
operators $a_{i}$ for a $\nu_e$ of momentum $p_i$ and $b_{i}$ for a $\nu_\tau$
of momentum $p_i$,  
where $i$ runs from unity to the number of single particle 
momenta.\footnote{Not 
necessarily equal to the number of particles, since in the initial 
state a  $\nu_e$ and a $\nu_\tau$ could both have the same momentum.}
Given an initial state of $N_1$ momenta occupied by $\nu_e$ and $N_2$ 
momenta occupied by $\nu_\tau$, we take the effective Hamiltonian that
implements the full set of reactions in Eq.~(\ref{reaction}) to be,
\begin{equation}
H_I^{(eff)}=\frac{1}{2} \frac{\sqrt{2} G_F}{V} \sum_{i \ne j}^{N_1+N_2} f_{ij}\,
[a^{\dagger}_{j} a_{i} b^{\dagger}_{i} b_{j} + \,a^{\dagger}_{i} a_{j} b^{\dagger}_{j} b_{i}]
\label{ham}
\end{equation}
where V is the volume of the system and the weight function $f_{ij}$ is of 
order unity. The sum extends over all of the $(N_1+N_2)$ momentum states of 
the system that are initially occupied by either flavor of neutrino. 
We have omitted the terms in the Hamiltonian corresponding to the 
processes, $\nu_e(p_i)+ \nu_\tau(p_j) \rightarrow \nu_e(p_i)+ \nu_\tau(p_j)$,
which do not contribute to our effects.  Including such terms, that is to 
say, a diagonal contribution to the Hamiltonian,
alters the wavefunction of the system only by an irrelevant overall phase.
Since the energies of the basic set of unperturbed states are exactly the same,
the time evolution of the system will be entirely determined by this 
interaction Hamiltonian.

In a realistic problem the form of $f_{ij}$ will depend on the circumstances of
the application. For example, if we had started with the complete form of the
neutral current, neutrino-neutrino interactions, then  the matrix elements of
the Dirac matrices in the V,A structure dictate a factor of  $[1-\cos(\theta
_{p_i,p_j})]$, where the angle is that between the two momentum vectors 
labeling
the states.  For simplicity, we have taken $f_{ij}=1$ $ \forall i,j$ for some 
of our analytic estimates.  In the numerical calculations we have taken a 
distribution of values in the range $0.5-1$, given by
\begin{equation}
f_{ij} = 0.5 + {0.5\over N_1+N_2} \left|i-j\right|.
\label{f}
\end{equation}

\section{N+N}
In our first example we shall take a set of $2N$ momentum states to be
occupied half by $\nu_e$'s and half by $\nu_\tau$'s.  For the initial state we
take the first (bottom) $N$ states to be filled by $ \nu_{\tau}$'s and the
last (top) $N$ states to be filled by $\nu_e$'s. We refer to this state as 
$|\Psi_0 \rangle$. Explicitly,
\begin{eqnarray}
\label{ic}
| \Psi_0 \rangle = | \nu_e(p_1) \ldots \nu_e(p_i) \ldots \nu_e(p_{N}) 
\nu_\tau(p_{N+1})
\ldots \nu_\tau(p_{j}) \ldots \nu_\tau(p_{2N})\rangle\,.
\label{thestate}
\end{eqnarray}
The interactions given in Eq.(\ref{reaction}) exchange energy and momentum
between particles of the two different species.  An example of such a process is the 
interchange of our initial state with 
\begin{eqnarray}
|\nu_e(p_1)  \ldots  \nu_\tau(p_i) \ldots \nu_e(p_{N}) 
\nu_\tau(p_{N+1}) \ldots \nu_e(p_{j}) \ldots \nu_\tau(p_{2N})\rangle\,.
\label{changedstate}
\end{eqnarray}
The total set of states of the complete system that we have
to deal with, $| \Psi_\alpha\rangle$, are the $n_s=(2N)! / (N!)^2$ distinct
states in which the flavor indices  in the initial state are permuted within
the defined subset indexed by $p_i$. We shall adopt an ordering of these states
such that  in each of the first $n_s/2$ states in the list, the top state in
the single particle list is occupied by $\nu_e$.  
Having taken all of the N 
of the $\nu_e$'s on the top, in the initial state, we wish to estimate the time for
``equilibration", in the sense of the $\nu_e$ 's being more or less distributed
with 50\% in the bottom set of states. 

In the $n_s$ dimensional space spanned
by our basis states, $|\Psi_\alpha \rangle$, the effective Hamiltonian matrix
as determined from Eq.(\ref{ham}) has $N^2$ off-diagonal elements in each 
row (or column); we denote this matrix by ${\bf M}$ . Each of the $n_s N^2$
off-diagonal elements of ${\bf M}$ is one of the 2N(2N-1)/2 values of the
function $f_{ij}$ of Eq.~(\ref{ham}).

Taking $f_{ij}=1$, we have,
\begin{equation}
\langle \Psi_\alpha | {\rm \bf  M}^2 
| \Psi_\alpha\rangle ={2 N^2 G_F^2\over V^2} 
\label{h2}
\end{equation}
for each one of the basis states $| \Psi_\alpha\rangle$. 
The square root of the average squared energy
of the $n_s$ eigenstates is thus given by
\begin{equation}
E_{av}=\Bigr[ n_s^{-1}Tr[{\bf M}^2] \Bigr ]^{1/2}= \frac{1}{2} 
\sqrt{2} G_F\, \rho
\label{eigen}
\end{equation}
where the trace operates in our space of $n_s$ states, and we 
have substituted the total $(\nu_e + \nu_\tau)$ number-density, 
$\rho$, for $2N/V$.
We denote the
corresponding set of eigenstates of $M$ by $|\Psi_E\rangle$, where the $E$'s stand 
for the $n_s$ eigenvalues.
The state of the system at time $T$, which we denote by $|\Psi(T)\rangle$ 
is given by
\begin{equation}
|\Psi (T)\rangle= \sum _E^{n_s} |\Psi _E\rangle \langle\Psi_E |
\Psi_0\rangle e^{-iET},
\label{evol}
\end{equation}
From this we can compute, for example, the probability $P$ (persistence) that 
a particular one of the top $n_s/2$ states, which was occupied at time T=0 
by a $\nu_e$, is occupied at time T by a $\nu_\tau $. For example, with the
$|\Psi_0\rangle$ and the ordering of the $n_s$ states as described above, 
we ask for the probability of continuing to have a $\nu_e$ in the very 
top state, obtaining
\begin{equation}
P= \sum_{\alpha  = 1}^{n_s/2}|\langle\Psi_\alpha   
 | \Psi (T) \rangle|^2,
\label{prob}
\end{equation}  
where the sum is over all the $n_s/2$ states that have a $\nu_e$ on top.

Of course at the moment we have said nothing about the coefficients
$ \langle\Psi_\alpha | \Psi_E \rangle$ that enter in 
Eq.~(\ref{prob}) and Eq.~(\ref{evol}). We
anticipate that an eigenvector of a matrix such as $M$ typically projects
significantly onto many of the basis states $|\Psi_\alpha \rangle$.  
Our numerical tests appear to sustain this conclusion, which we will not 
quantify further in this note.
Then we conjecture that the time in which $P$ becomes significantly
less than unity on the average, i.e. the mixing time, is determined by the
magnitude 
of a typical
eigenvalue. From  Eq.~(\ref{eigen}) we already found that the root mean 
square of an eigenvalue is of order $\rho\,G_F$. If a macroscopic fraction of the
eigenvalues, that is, a number $\lambda \, n_s$ of them, are  of this
order,\footnote{Of course, if essentially all of the strength were concentrated
in a number of  eigenvalues which did not grow as $n_s$, then we would not
obtain macroscopic effects.} then the effective mixing time should be of order
$(\rho\,G_F)^{-1}$, in view of the relations Eq.(\ref{evol}) and
Eq.(\ref{prob}). This is
the same as the typical time scale for what the ``index of refraction" or
``forward-scattering" effects would be in the case of in which we replaced the
neutrino density by an electron density of the same magnitude, a time scale
much shorter  than any effect of nonforward scatterings (which scale as 
$G_F^{-2}$.)

The distribution of the eigenvalues can be determined analytically in the case in which 
$f_{ij}=1$.  Consider a system in which the number of $\nu_e$ and $\nu_\tau$  
is given by $N_1$ and $N_2$ respectively.  We shall assume, without loss of
generality, that $N_1\leq N_2$.  The distribution of eigenvalues have a 
pattern, from which we can observe that there are 
$N_1 + 1$ distinct eigenvalues 
\begin{equation}
E_i = \frac{\sqrt{2}G_F}{V}
[(N_1-i)(N_2-i)-i], \,\,\,\,\,\,\,\,\,\,\,\,\ i=0,1 \ldots N_1,
\end{equation}
with degeneracies given by
\begin{equation}
D_i={  (N_1+N_2)!  \over i!(N_1+N_2-i)!}-{(N_1+N_2)!\over (i-1)!(N_1+N_2-(i-1))!}. 
\end{equation}

\begin{figure}[ht]
\begin{center}
\epsfig{file=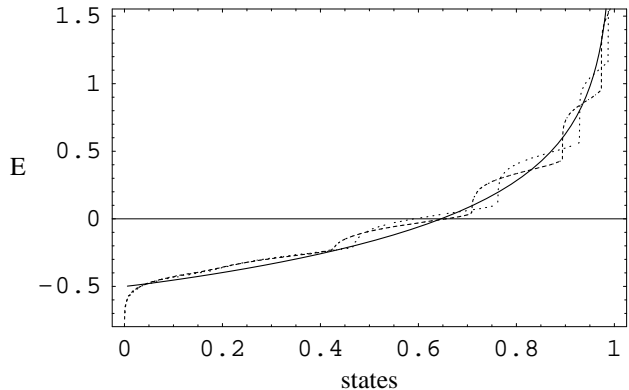,width=3.25in}
\caption{\label{eval}
The distribution of energy eigenvalues for a system of $N$ $\nu_e$'s and 
$N$ $\nu_\tau$'s, where the energy is in units of $\sqrt{2}G_F \rho$.   
A point on the curve represents the fraction of states with energy below $E$.
The heavy curve is the limiting result for large 2$N$ for the case $f_{ij}=1$. 
The heavier and lighter dashed curves are the results for 
2$N$=14 and 2$N$=12 respectively, with $f_{ij}$ given by Eq.~(\ref{f}), 
but multiplied by a factor of 4/3 so that the average coupling strengths 
in the comparisons are the same.}
\end{center}
\end{figure}

In Fig.~\ref{eval} we plot the distribution of these eigenvalues for the present case of 
$N$ each of $\nu_e$ and $\nu_\tau$, in the limit of large N.  On the same figure
we show the distributions obtained numerically, in the case that $f_{ij}$ is given
by Eq.~(\ref{f}), for the cases $2N=12$ and $2N=14$.
Since the matrix $M$ is traceless, the eigenvalues sum to zero. 
One can see from Fig.~(\ref{eval}) that the bulk of the eigenvalues 
(say, $>95\%$) lie in the range 
\begin{equation}
- \frac{1}{2}  \leq \frac{E_i}{2N} \frac{V}{\sqrt{2}G_F} \leq + 1,
\end{equation}
and thus the typical energy difference between a pair of eigenstates 
will be of order $(\sqrt{2}G_F/V)(2N) \equiv \sqrt{2}G_F \rho$.  
The qualitative features of the eigenvalue distribution, are not affected 
by taking a distribution of $f_{ij}$'s.


\begin{figure}[ht]
\begin{center}
\epsfig{file=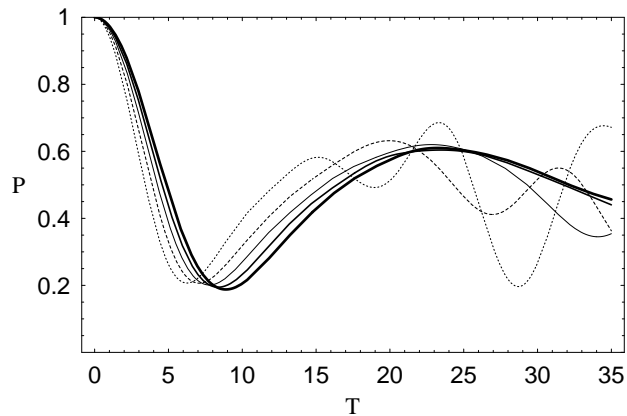,width=3.25in}
\caption{\label{nn}
The mixing parameter, or single level persistence probability, $P$,
as defined in text, as a function of time, for a system of $N$ $\nu_e$
and $N$ $\nu_\tau$.  The curves correspond to the number of particles:
2$N$=6, 8, 10, 12 14, where the heaviest curves are for highest N. The unit
of time is $(\sqrt{2} G_F\, \rho)^{-1}$.}
\end{center}
\end{figure}

We also make a limited computational check on our heuristic
estimates of the effects on rates, by directly solving the 
Schrodinger equation,
beginning with the above initial state, Eq.~(\ref{ic}), and determining the 
probability that a
particular one of the $\nu_e$'s in one of the 
upper states has been replaced at a later time by a $\nu_\tau$. 
In the case $2N=14$ we have to solve $n_s=14!/(7!)^2=3432$ coupled linear 
differential
equations, which is our computational limit. The results of these calculations
are shown in Fig.~\ref{nn} for a case in which the coupling function 
$f_{ij}$ is given by Eq.~(\ref{f}). We see that the curves show very 
similar behavior, and except
for the smallest case with 6 particles, the times elapsed to the first minimum
are very close to each other. The results appear to sustain the analysis, given
above, for a mixing rate that is independent of particle number for fixed
density. Note, however, that the locations for, say the first peaks in the
respective curves move to greater time as N is increased, albeit at a slower
and slower rate. One cannot absolutely conclude from these data that there is a
limiting point for large N and fixed density. Provisionally, taking the mixing
time to 
be that in which the
curves cross in Fig.~\ref{nn}, we find 
$T_{\rm mix} \approx (\sqrt{2}G_F\, \rho)^{-1}$ in the limit of an infinite system. 
Unfortunately, a perturbation theoretic approach is valid only for small times
(smaller and smaller times as the particle number is increased) and cannot shed
any light on the outcome.

\section{1+(2N-1)}

\begin{figure}[ht]
\begin{center}
\epsfig{file=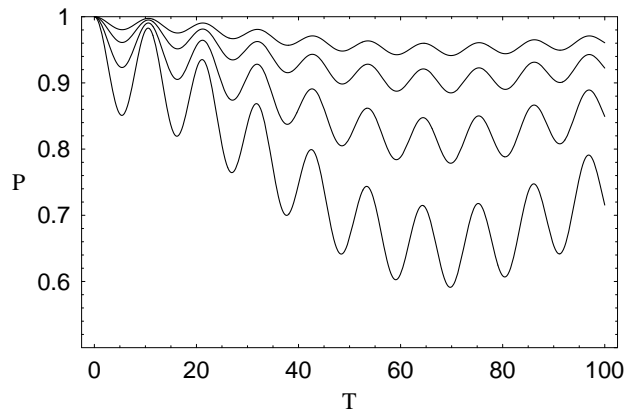,width=3.25in}
\caption{\label{1n}
The same as Fig.2 except for the case of an initial
system state with only one particle state occupied by a $\nu_e$ and the
others occupied by $\nu_\tau$'s, with P the probability that this the
$\nu_e$ is in its initial state. We compare the cases $2N$=30, 60, 120, 240
where the slower evolution corresponds to the higher number of
particles.}
\end{center}
\end{figure}

Next we look at a problem that is so closely related to the above one that one
might (erroneously) guess that the behavior is similar.  We start with the same
Hamiltonian Eq.~(\ref{ham}), and set of $2N$ participating states but now take
an initial condition with only the top state filled with a $\nu_e$ and all the
remaining $(2N-1)$ single-particle states filled with $\nu_\tau$'s. The effective
subspace of system states that are connected together is now $n_s=2N$
dimensional, and is described by the location of the single $\nu_e$. But now,
in the analogue of Eq.~(\ref{h2}), for the case of $f_{ij}=1$ we find that the
average squared eigenvalue is 
$(2N)^{-1} {\rm Tr } {\bf M}^2=(2N-1)(\sqrt{2}G_F/V)^2$, in contrast
to the value, $N^2$, obtained in the $(N,N)$ case.  Furthermore, there is a
single state, with eigenvalue $(2N-1)$, while the  remaining $(2N-1)$ states
all have eigenvalue $-1$. Denoting the state with large eigenvalue by $\Psi_S$,
we note that $|\langle \Psi_S|\Psi_0 \rangle|^2=N^{-1}$ so that the effects on
$P$ of mixing with this state are of order $N^{-1}$; and since there are no
energy differences among the remaining $N-1$ eigenstates, the total effect (at
the order $G_F$ level) will vanish in the limit of large N.  We have confirmed
numerically that the introduction of a scatter into the coupling constants,
using Eq.~(\ref{f}), does not change these conclusions qualitatively.  The
results for persistence versus time in this case are shown in Fig.~\ref{1n},
for the case of several values of $N$.  They clearly show the $N^{-1}$ behavior
in the short and intermediate time regions.

\section{Discussion}

Thus far in our discussion, we have used ``$\nu_e$" and ``$\nu_\tau$" merely to
label our states, and we have given no hint as to how the above behaviors could
make a difference to observable results in the neutrino world. In fact there
are two situations that have been discussed in the literature in which
neutrino-neutrino scattering is thought to play a role, and in which the size
of the parameters is exactly such as to make effects with our time scale 
$(\sqrt{2} G_F \rho)^{-1}$ of possible importance, and indeed as important 
as the effects
noted in the literature. The first is in early universe scenarios in which one
has assumed some degree of neutrino degeneracy \cite{earlyu}.  
The second is in the region
just above the neutrino-sphere in the supernova process, 
where differences in the energy spectra for different neutrino flavors
play a key role in whether or not there is an efficient R process for
nucleosynthesis \cite{sn}.
Another example where neutrino-neutrino forward scattering is relevant,
is the flavor conversion of high energy neutrinos from astrophysical 
sources as they propagate through the relic neutrino background \cite{lunardini}.
We emphasize again
that up to this point we have not included the neutrino oscillation physics
that drives the results in these works. We intend to address the combination of
the two themes  in a subsequent publication. But we can comment on what we
believe to be the insufficiency of the previous works on the subject. What we
have computed above, the probability that a specific state, say $p_i$, that was
originally occupied by a $\nu_e$ (in the initial state for the complete system
$|\Psi_0\rangle$) is still occupied by a $\nu_e$ at time $t$, has a simple 
expression in terms of the Heisenberg picture operator, $a_{i}(t)$,

\begin{equation}
P=\langle \Psi_0 | a^\dagger_{i}(t) a_{i}(t)| \Psi_0 \rangle,
\label{heis}
\end{equation}
where the Heisenberg operators are chosen to coincide with the Schrodinger
operators that were introduced
in Eq.~(\ref{ham}) at $t=0$. Now the operator in Eq.~(\ref{heis}) is a 
neutrino density operator for the
particular state $p_i$. If we write an equation of motion for this operator, by
taking the commutator with
the Hamiltonian Eq.~(\ref{ham}), the right hand side is a quartic in 
the operators $a^\dagger , a , b^\dagger , b$,
and there is a sum over one index $j$. When we encounter, say, a term like
\begin{equation}
 \sum_j a^\dagger_{j}(t) b_{j}(t) b^\dagger_{i}(t) a_{i}(t)
\label{sayaterm}
\end{equation}
we would \underline{like} to be able to replace  $\sum_j (a^\dagger_{j}
b_{j} )$ by a density matrix  element, off-diagonal in the flavor 
space,
which represents the entire average state of the medium. Then the multiplying
operator $b^\dagger_{i} a_{i}$ would be a corresponding operator but for 
the single mode $i$. Indeed, this is exactly the assumption that yields the 
non-linear
terms in the equations for the density matrices derived in Ref.~\cite{raff} and 
used in Ref.~\cite{earlyu,sn}. 

As noted
above, with our assumed flavor-diagonal initial conditions, and in the absence
of  neutrino-mixing in the  Hamiltonian, such terms do not create any
effect. Our effects are exactly due to the fact that the replacement of the
four-operator product by a product of two expectation values, as sketched
above, is not justified. Even if we were to assume that
a kind of factorized ansatz would pick up the leading terms, which we believe
is highly unlikely, we would want add the results of pairing, say, 
$a^\dagger_{j} a_{i}$ in Eq.~(\ref{sayaterm}). That is to say, we would be
driven to consider density matrices that are
off-diagonal in momentum space, as well as in flavor space.

To summarize, we have evidence for a new, macroscopic quantum effect that could
change the outcome of calculations in which $\nu$-$\nu$ scattering matters. 
The caveats that must be added are:

(i) Accepting the earlier conclusions for the model defined by the 
Hamiltonian Eq.~(\ref{ham}), we must go back and ask whether we 
have defined the correct problem.
All those modes that we left out, when we truncated our Hamiltonian to 
momentum states that were initially occupied by one species or the other, 
can they make a difference? 
These additional modes would be populated via non-forward scattering
interactions, which enter the problem at the level of cross-section ($\times$
density), that is, on time scales of order  $(G_F^2 \rho\,\omega^2)^{-1}$.  
Our expectation is that they will not affect our results on the much shorter time
scales we consider (of order  $(G_F \rho)^{-1}$.

(ii) Our calculations pertain directly to plane waves in a volume $V$. Ordinary neutrino 
oscillation theory has come under fire repeatedly from authors that suspect
that preparation-of-state considerations bring the idealized, plane-wave
picture into question. These criticisms have been successfully answered more
than once \cite{lipkin} for the case of single-particle oscillation,
but they could surface again in our present context.

At the present time, we are cautious about claiming that our results
will be important to the neutrino physics either in the early universe or
in the supernova. These are, however, the two most obvious applications in
which the neutrino number density is high enough for neutrino-neutrino
scattering to be important.  
We further note that the flavor-energy correlation is critical to understanding
the physics just outside the supernova neutrinosphere. 
Of course, we would need to include neutrino mass 
and mixing in the model
in order to address a realistic situation. Superficially, we can note that the 
energies attributable to mass effects are small compared to the inverse time 
associated with our processes under the conditions that prevail just outside
the neutrinosphere of a supernova. But more analysis is required before
reaching a conclusion in that context. 
A negative feature
of our work is that in contrast to being able to use the quite simple equations
for the density matrix posited by Ref.~\cite{raff}, we appear to be doomed to
treating the full complexity of the many-body physics that arises in these systems.

\section{An example with a finite number of discrete states}
We demystify the physics of much of the above, to some degree,
by thinking of a fairly large but finite set of discrete quantum 
subsystems
with interactions of comparable strength between each pair. In 
addition to being an interesting exercise this also is relevant to
applications in other arenas than the world of neutrinos. We consider
a set of N spins associated with one set of fixed site labels, say red,
and another set of N spins associated with a second set of labels, say 
blue.
The interaction exchanges all pairs of spins with comparable strength, or order
$g$, for each pair.
Let us now start with all reds up 
and 
all blues down, and consider the short time evolution of this state. The 
probability that the spin on a particular red site becomes down in a very short time 
$t$
is of order $h =g^2 N t^2$. We might think that this perturbation 
calculation stays
more or less valid until such a time that $h$ is, say, $0.1$. But if we go 
back
through the same mechanics that we used for our continuum problem, we
see that is not be the case; the system will mix in a time $T$
such that  $g N T=1$\footnote{Note that in this case we do not 
redefine the
coupling constant with increasing N; there is no volume or density 
involved
in the models. Thus the eigenvalues grow as N.}, at which time the 
above perturbative estimate of the 
amount of mixing is only $h=1/N$.

Next consider the case in which the first red spin is up, and every other  red
and blue spin is down. In complete correspondence with the (1, 2N-1)  neutrino
case, we find that the above perturbative estimate of the mixing time, for the first
red  spin to become appreciably blue, is correct. The mixing time for this
single  state is longer than that of the first example by a factor of $N$, for
large $N$. 

What makes these cases so different from each other? In both cases
the spin at the first red site interacts with each of the N blue spins. But in
the second  example, the blue spins themselves interact only with that one red
site, at the turn-on time, while in the  first example every blue spin is being
affected by every red spin from the beginning. The difference lies not in  the
number of interactions that the red spin sees, but in the entanglements of the
states that it interacts with. Since in the first case we get faster evolution
than we  would have  expected from the perturbation estimate we classify the
effect as a  ``speed-up".  And the reason is clearly the multiparticle
entanglement within the system. For completeness, we should note that even in the second
example, the blue sites, with which the spin on the distinguished red site interacts, do develop
mutual entanglement, but with a much slower initial rate than in the first example.
In effect, they see each other only through their mutual coupling to the distinguished
red site. 

\section{Conclusion}
We have examined a ``speed-up'' of evolution through entanglement, both in a
discrete system of spins and in a neutrino model with a continuum of states.
Though the practical meaning of ``speed-up" is slightly different in the two
cases, the formal source of the effect is the same. 

The reader might ask the question: ``Are the authors using the word
`entanglement' in some precise mathematical sense, or are they just using it to say
the system gets complicated?" 
We contend that it is the former, in the sense that the state of the system
cannot, in general, be factorized into a product of single particle states.  
Note, however, that the quantification of multiparticle entanglement is not a
concept that that has been precisely defined in the literature. 
Definitions of two-state entanglement, however, have received much
attention. The reader can consult Ref. \cite{entangle} for a demonstration of
how in the simplest relevant ordinary Schrodinger example, with two particles
in a double well, entanglement in a precise mathematical sense is spontaneously 
generated if the two particles interact with each other. 

Returning to the neutrino example, the first order
effects that mixed the states had an inverse time-scale $\sqrt{2} G_F \rho$; 
this is
the ``speeded-up" rate.  The  usual inverse time-scale for (non-forward) 
scattering effects is of order $G_F^2 \rho\,\omega^2$ where $\omega$ is of the 
order of  the particle energies and much, much slower, than the ``speeded-up"
rate when $\omega$  is of order an MeV.
Comparing again with the calculation with only one $\nu_e$ in a sea of
$\nu_\tau $'s, where there is no speed-up,  the difference is 
attributable to the degree of entanglement.

The inclusion of entanglement requires abandoning the single-particle
description of the system, to include the full complexity of the many-body 
physics involved.
Single-body descriptions do predict significant neutrino-neutrino forward
scattering effects \cite{pantaleone,raff}, which, however, are absent for the
flavor diagonal initial states we consider here.  The results of our many-body
calculation may thus have interesting consequences in situations where 
$\nu$-$\nu$ scattering is important.

\subsection*{Acknowledgements}
\noindent
We thank John Beacom, Boris Kayser, Cecilia Lunardini and 
Doug Scalapino for helpful conversations.  
The work reported here began in interchanges at the neutrino workshop at the 
Kavli Institute for Theoretical Physics at UCSB, supported by the 
National Science Foundation under Grant No. PHY99-07949.
It was motivated, in part, by a talk given at KITP by Alex Friedland, which is
available on the web at 
http://online.kitp.ucsb.edu/online/neutrinos03/friedland/.
NFB was supported by Fermilab (operated by URA under DOE contract 
DE-AC02-76CH03000) and by NASA grant NAG5-10842. 
The computations were done using Mathematica.

{\bf NOTE ADDED}

After our paper appeared, Friedland and Lunardini \cite{FL2} analyzed the
totally symmetric case, where all the coefficients $f_{ij}=1$, and have
shown that the mixing times (for couplings scaled to 1/[particle number])
grow as $\sqrt{N}$, for large N, in contradiction to our conjecture.
However, this analytic solution cannot be readily extended to the case of
unequal $f_{ij}$, so the results of \cite{FL2} are perhaps inconclusive in
this case. As explained in the text, the simulations presented in
Fig. 2 are for the case of scattered coupling strengths. These
simulations are not adequate for distinguishing $\sqrt{N}$ limiting
behavior from constant limiting behavior, due to the restrictions on N
imposed by computational resources.

We have since applied the techniques of \cite{FL2} to some cases of
differing coupling schemes and initial conditions, but still with
sufficient symmetry to do simulations for much larger values of N than
previously, and we find evidence for fast evolution in some of these more
generalised systems.

\end{document}